\documentclass[12pt]{iopart}

 \usepackage{iopams}
  \usepackage{amsthm}
\newfont{\myeu}{eurm10 at 12 pt}
\newfont{\bfrak}{eufb10 at 12 pt}

\def\ba{{\bar a}}
\def\bb{{\bar b}}
\def\bG{{\bar G}}
\def\bI{{\bar I}}
\def\bJ{{\bar J}}
\def\bK{{\bar K}}
\def\bN{{\bar N}}

\def\vp{\vphantom{P}}
\def\sfactor#1#2{\Bigg[\begin{array}{@{}c@{}}#1\\#2\end{array}\Bigg]}
\font\my=cmr12 at 14pt
\def\myphi{\hbox{\my\char'010}}
\def\hypp#1#2#3#4#5{{}_{#1}
\myphi_{#2}\left[{#3\atop\phantom{\omega,}#4};#5\right]}

\def\sfactor#1#2{\Bigg[\begin{array}{@{}c@{}}#1\\#2\end{array}\Bigg]}
\def\bino#1#2{\Bigg(\begin{array}{@{}c@{}}#1\\#2\end{array}\Bigg)}

\def\halfs{{\scriptstyle{\frac 12}}}
\begin{document}

\title [Identities in the Superintegrable Chiral Potts Model ]%
{Identities in the Superintegrable Chiral Potts Model}
\author{Helen Au-Yang$^{1,2}$ and Jacques H H Perk$^{1,2}$%
\footnote{Supported in part by the National Science Foundation
under grant PHY-07-58139 and by the Australian Research Council
under Project ID: LX0989627.}}
\address{$^1$ Department of Physics, Oklahoma State University,
145 Physical Sciences, Stillwater, OK 74078-3072, USA%
\footnote{Permanent address.}}
\address{$^2$ Centre for Mathematics and
its Applications \& Department of Theoretical Physics,
Australian National University, Canberra, ACT 2600, Australia}
\ead{\mailto{perk@okstate.edu}, \mailto{helenperk@yahoo.com}}

\begin{abstract}
We present proofs for a number of identities that are needed to study
the superintegrable chiral Potts model in the $Q\ne0$ sector.
\end{abstract}

\vspace{2pc}

\section{Introduction}

The integrable $N$-state chiral Potts model is defined through a
solution of the star-triangle equations parametrized by a Fermat-type
curve of genus greater than one \cite{AMPTY,BPA}. In spite of this
difficulty, eigenvalues of the transfer matrix have been obtained using
functional equations \cite{BBP,BS}. There are now detailed results for
free energies, interfacial tensions and excitation spectra \cite{BaxFE,BaxIf1,BaxIf2,AMP,MR}. As far as the correlation functions
are concerned, not much is known beyond the order parameter
\cite{AMPT,BaxO} and some limited partial results on the leading
asymptotic behavior of pair correlation functions from conformal field
theory and finite-size corrections \cite{AM}. Further progress requires
the knowledge of eigenvectors of the transfer matrix and form factors.

In order to proceed two recent approaches have been proposed, both
starting by studying eigenvectors of the $\tau_2$ model \cite{BBP,BS}.
The first approach \cite{Lisovyy,Iorgov,GIPS,IST,GIPST,GIPST2,GIPS2}
begins with the $B_n(\lambda)$ element of the monodromy matrix of
the $\tau_2$ model, diagonalizing it for fixed boundary conditions.
Using the $B_n(\lambda)$ eigenvectors further results are found for
the $\tau_2$ transfer matrix and commuting spin chain Hamiltonians,
with the most explicit results for the finite-size $N=2$ Ising case.
The other (our) approach is to start from the superintegrable $\tau_2$
model with periodic boundary conditions in order to find eigenvectors
of the $N$-state chiral Potts model.

In our two previous papers~\cite{APsu1,APsu2}\footnote{All equations
in~\cite{APsu1} or~\cite{APsu2} are denoted here by prefacing I or II
to their equation numbers.} we have expressed the transfer matrices
of the superintegrable chiral Potts model for the $Q=0$ ground state
sector---with $Q$ the spin shift quantum number and length $L$ a
multiple of the number $N$ of states for each spin---in terms of
generators of simple $\frak{sl}_2$ algebras, which
generate the corresponding degenerate eigenspace of $\tau_2(t_q)$.
The $\tau_2(t_q)$ used in our papers are directly obtained from the
chiral Potts model and are different from those of Nishino and Deguchi
\cite{NiDe1, NiDe2}. However, as they are related \cite{NiDe2, APsu4},
one can easily follow the procedure of  \cite{NiDe1} to obtain
the corresponding loop algebra $L({\mathfrak{sl}}_2)$, which can be
decomposed into $r=(N\!-\!1)L/N$ simple ${\mathfrak{sl}}_2$ algebras
\cite{APsu2}. Thus we found $2^r$ chiral Potts eigenvectors for $Q=0$.

For the $Q\ne 0$ cases, the situation is more difficult, and not much
is known about the eigenvectors in the published literature%
\footnote{Tarasov \cite{Tarasov} has written the Bethe Ansatz
eigenvectors of the special superintegrable $\tau_2$ transfer matrix
with vertical rapidities $p\equiv p'$ for general $Q$. However, due to
the high level of degeneracy of the corresponding eigenvalues, he
could not give the related chiral Potts eigenvectors.}, except
that some related investigation on the six-vertex model at roots of
unity has been done in \cite{DFM}. As can be seen from e.g.\ (I.47) in
\cite{APsu1}, and (45) of the follow-up \cite{APsu4} of this paper,
there are different ways to construct the eigenvectors of the
degenerate eigenspaces of $\tau_2(t_q)$. It is highly nontrivial,
therefore, to determine the generators of the loop algebra
$L({\mathfrak{sl}}_2)$.  

We propose that (II.66) and (II.68) for $Q=0$ can be generalized
to $Q\ne0$ cases as
\begin{eqnarray}
\langle \Omega|{\bf E}_{m,Q}^{-}=-(\beta^Q_{m,0}/\Lambda^Q_{0})
\sum_{ {\{0\le n_j\le N-1\}}\atop{n_1+\cdots+n_L=N}}\langle\{n_j\}|\,
\omega^{\sum_{j}j n_j}\bar{G}^{\vp}_Q(\{n_j\},z^{\vp}_{m,Q}),
\label{ome2}\\
{\bf E}_{k,Q}^{+}|\Omega\rangle=(\beta^Q_{k,0}/\Lambda^Q_{0})
z^{\vp}_{k,Q}\sum_{ {\{0\le n_j\le N-1\}}\atop{n_1+\cdots+n_L=N}}
\omega^{-\sum_{j}j n_j}
{G}^{\vp}_Q(\{n_j\},z^{\vp}_{k,Q})|\{n_j\}\rangle.
\label{epo2}
\end{eqnarray}
where $z^{\vp}_{k,Q}$, ${G}^{\vp}_Q$ and ${\bar G}^{\vp}_Q$ shall be
defined later in the paper. If such a generalization is valid, or,
in other words, if the ${\bf E}_{m,Q}^{\pm}$ are indeed the generators
of ${\mathfrak{sl}}_2$ algebras, then it is necessary that
\begin{equation}
\langle\Omega|{\bf E}_{k,Q}^{-}{\bf E}_{m,Q}^+|\Omega\rangle
=-\delta_{k,m}\langle\Omega|{\bf H}_{k}^{Q}|\Omega\rangle
=\delta_{k,m}.
\label{ortho0}
\end{equation}
In this paper we shall prove this orthogonality relation, which is
used in \cite{APsu4} to find generators of an $L({\mathfrak{sl}}_2)$
loop algebra for $Q\ne0$. For $Q=0$, this proof is not necessary, but
can be seen as added confirmation that the decomposition of the loop
algebra to ${\mathfrak{sl}}_2$'s is correct.

The results of this paper are used in \cite{APsu4} for the case of the
superintegrable chiral Potts model with periodic boundary conditions,
$L$ a multiple of $N$ and $Q\ne0$. It should be remarked that the
results of this paper also apply to cases with $L$ not a multiple
of $N$, but in the thermodynamic limit $L\to\infty$ one can ignore
these cases. Finally, as the identities in this work may also be
related to combinatorics, this paper will be presented with details
using a self-contained presentation.

\section{Drinfeld Polynomials and Other Definitions}

In (I.40), we have considered the polynomial
\begin{equation}
{\cal Q}(t)=\prod_{j=1}^L\;\sum_{n_j=0}^{N-1}t^{n_j}=
\frac{(1-t^N)^L}{(1-t)^L}=\sum_{m=0}^{L(N-1)} c_m t^m,
\label{Q}\end{equation}
whose coefficient $c_m$ is the number of $L$-dimensional vectors
$\{n_j\}=\{n_1,\cdots,n_L\}$ with elements that are integers
$n_j=0,\dots, N-1$ satisfying the condition  $n_1+\cdots+n_L=m$.
In this paper, $Q$ denotes a nonnegative integer less than $N$.
The Drinfeld polynomial $P_Q(z)$ is defined in (I.11) and can be
expressed in terms of the $c_m$ as
\begin{eqnarray}
&P_Q(z)&=c^{\vphantom{p}}_Q+c^{\vphantom{p}}_{N+Q}\,z+\cdots
+c^{\vphantom{p}}_{m_QN+Q}\,z^{m_Q}\nonumber\\
&&=N^{-1}t^{-Q}\sum_{a=0}^{N-1}\omega^{-Qa}{\cal Q}(t\omega^a)
=N^{-1}t^{-Q}\sum_{a=0}^{N-1}
\omega^{-Qa}\frac{(1-t^N)^L}{(1-\omega^a t)^L},
\label{DrinP}\end{eqnarray}
where
\begin{equation}
m_Q\equiv\lfloor(N-1)L/N-Q/N\rfloor,\quad
\omega\equiv\rme^{2\pi\rmi/N}, \quad z\equiv t^N.
\end{equation}

We have also defined in (II.60) and (II.61) the $L$-fold sums 
 \begin{eqnarray}
&&K_m(\{n_j\})=\sum_{ {\{0\le n'_j\le N-1\}}\atop{n'_1+\cdots+n'_L=m} }
\prod_{j=1}^L\sfactor{n_j+n'_j}{n'_j}
\omega^{n'_j N_j},\quad N_j=\sum_{\ell=1}^{j-1}n_\ell\,, \label{sumK}\\
&&\bK_m(\{n_j\})=\sum_{{\{0\le n'_j\le N-1\}}\atop{n'_1+\cdots+n'_L=m}}
\prod_{j=1}^L\sfactor{n_j+n'_j}{n'_j}
\omega^{n'_j\bN_j},\quad \bN_j=\sum_{\ell=j+1}^{L}n_\ell\,,
\label{sumbK}\end{eqnarray}
where the $\omega$-binomial coefficients \cite{AnAsRoy} are defined by
 \begin{eqnarray}
&&[n]\equiv 1+\cdots+\omega^{n-1},\qquad [n]!\equiv[n]\,[n-1]\cdots[1],
\nonumber\\
&&\sfactor{n}{n'}\equiv\frac{[n]!}{[n']!\,[n\!-\!n']!}=
 \frac{(\omega^{1+n-n'};\omega)_{n'}}{(\omega;\omega)_{n'}},
  \quad (x;\omega)_n\equiv\prod_{j=1}^n(1-x\omega^{j-1}).
\label{binomial}\end{eqnarray}
We remove the constraint $n'_1+\cdots+n'_L=m$ in (\ref{sumK}) by
inserting $\prod t^{n'_j}$ within the $L$-fold summation and summing
over $m$ from 0 to $(N-1)L$. Next we use
\begin{eqnarray}
&&\sfactor{n+r}{r}=(-1)^r\omega^{nr+r(r+1)/2}
\sfactor{N\!-\!1\!-\!n}{r},\label{id1}\\
&&\sum_{r=0}^s \sfactor{s}{r}(-1)^r\omega^{r(r-1)/2}x^r=(x;\omega)_s,
\label{id1a}\end{eqnarray}
see  (10.2.2c) in \cite{AnAsRoy}, to perform each of the $L$ sums,
noting $[{s\atop r}]=0$ for $r>s$. After using
\begin{eqnarray}
&\fl(x\omega^{s+1};\omega)_{N-1-s}=\frac{(1-x^N)}{(x;\omega)_{s+1}}
=\frac{(1-x^N)}{(1-x)(x\omega;\omega)_s},\qquad
&(x;\omega)_s\;(\omega^s x;\omega)_{r}=(x;\omega)_{r+s},\nonumber\\
&\fl \prod_{j=1}^L\,(t\omega^{1+N_j};\omega)_{n_j}
=(t;\omega)_{n_1+\cdots+n_L}=(t;\omega)_{N_{L+1}},
&N_1=0,\quad N_j+n_j=N_{j+1},
\label{id2}\end{eqnarray}
we obtain the resulting  generating function for
$N_{L+1}=n_1+\cdots+n_L=k N$ as
\begin{equation}
g(\{n_j\},t)
=(1-t^N)^{L-k}\prod_{j=1}^L\,(1-t \omega^{N_j})\strut^{-1},
\label{g1}\end{equation}
which is also given in (II.62).
The condition $n'_1+\cdots+n'_L=m$ in (\ref{sumK}) means that
$K_m(\{n_j\})$ is the coefficients of $t^{m}$ 
in the expansion of $g(\{n_j\},t)$, i.e.,
\begin{equation}
g(\{n_j\},t)=\sum_{m=0}^{(N-1)L-k N}K_m(\{n_j\})t^m.
\label{g2}\end{equation}
In the same way, we can derive the generating function
of $\bK_m(\{n_j\})$. We end up with an equation like (\ref{g1})
where $N_j$ is replaced by $\bar N_j=kN-N_{j+1}$,
so that
\begin{equation}
{\bar g}(\{n_j\},t)=\sum_{m=0}^{(N-1)L-k N}\bK_m(\{n_j\})t^m=
g(\{n_j\},t)^*,
\label{bg2}\end{equation}
which is the complex conjugate of (\ref{g2}) when $t$ is real. 

In the special case $n_1\!=\!\cdots\!=\!n_L\!=\!0$, we have $k\!=\!0$,
and it can be seen from (\ref{sumK}) that $N_j\!=\!0$ for all $j$.
Comparing (\ref{g1}) with (\ref{Q}), we find
$g(\{0\},t)\!=\!{\bar g}(\{0\},t)\!=\!{\cal Q}(t)$.
This also means that $K_m(\{0\})\!=\!\bK_m(\{0\})\!=\!c_m$. 

In (II.63) and (II.64) we have defined the polynomials
\begin{eqnarray}
&&\fl G_Q(\{n_j\},z)
=N^{-1}t^{-Q}\sum_{a=0}^{N-1}\omega^{-Qa}g(\{n_j\},t\omega^a)
=\sum_{m=0}^{m_Q-k}K_{mN+Q}(\{n_j\})z^m,
\nonumber\\
&&\fl{\bG}_Q(\{n_j\},z)
=N^{-1}t^{-Q}\sum_{a=0}^{N-1}\omega^{-Qa}{\bar g}(\{n_j\},t\omega^a)
=\sum_{m=0}^{m_Q-k}\bK_{mN+Q}(\{n_j\})z^m.
\label{GbG1}\end{eqnarray}
The polynomials (\ref{GbG1}) correspond to (\ref{g2}) and (\ref{bg2})
in the same way as (\ref{DrinP}) relates to (\ref{Q}),
and for $Q=0$, they are related to the generators ${\bf E}_m^{\pm}$
of a direct sum of $\mathfrak{sl}_2$ algebras \cite{APsu2}. We shall
show that they satisfy the following orthogonality relation for all $Q$.
\goodbreak

\section{Main Theorem and Other Identities}

\newtheorem*{ort}{Theorem}
\begin{ort}
Let the roots of the Drinfeld polynomial $P_Q(z)$ given in (\ref{Q})
be denoted by $z_{j,Q}$, for $j=1,\cdots,m_Q$, i.e.
\begin{equation}
P_Q(z)=\sum_{m=0}^{m_Q}\Lambda^{Q}_{ m}z^m=\Lambda^{Q}_{ m_Q}\prod_{j=1}^{m_Q}(z-z_{j,Q}),\quad \Lambda^Q_m\equiv c^{\vp}_{mN+Q}.
\label{roots}\end{equation}
Then the polynomials (\ref{GbG1}) satisfy the orthogonality relation
\begin{equation}
\sum_{ {\{0\le n_j\le N-1\}}\atop{n_1+\cdots+n_L=N}}
\bG_Q(\{n_j\},z_{i,Q})\,G_Q(\{n_j\},z_{k,Q})
=-B_k\delta_{ik}\,,
\label{ortho}\end{equation}
where $B_k$ is a constant given by
\begin{equation}
B_k=z_{k,Q} (\Lambda^{Q}_{m_Q})^2\prod_{\ell\ne k}(z_{k,Q}-z_{\ell,Q })^2.
\label{conB}\end{equation}
\end{ort} 

When this orthogonality identity holds for any $Q$, then
(II.66)--(II.69) can be generalized to $Q\ne0$. Originally we used
Maple to check it for $N$ and $L$ small, and found that it indeed holds.
Here we shall present a proof by first proving a few lemmas.\goodbreak
\newtheorem*{Isum}{Definition}
\begin{Isum}
Let $\mu_j$ and $\lambda_j$ be integers satisfying
$0\le\mu_j,\lambda_j\le N-1$, for $j=1,\cdots,L$, and let
\begin{equation}
a_j=\sum_{\ell=1}^{j-1}\mu_\ell ,\quad
\ba_j=\sum_{\ell=j+1}^{L}\mu_\ell,\quad
b_j=\sum_{\ell=1}^{j-1}\lambda_\ell,\quad
\bb_j=\sum_{\ell=j+1}^{L}\lambda_\ell.
\label{ab}\end{equation}
Then $I_m(\{\mu_j\};\{\lambda_j\})$ is the $L$-fold sum depending
on $m$, $\{\mu_j\}$ and $\{\lambda_j\}$ and defined by
\begin{equation}
I_m(\{\mu_j\};\{\lambda_j\})\equiv
\sum_{ {\{0\le n_j\le N-1\}}\atop{n_1+\cdots+n_L=m}}
\prod_{j=1}^L\sfactor{\mu_j}{n_j}\sfactor{n_j+\lambda_j}{n_j}\omega^{n_j(a_j-N_j)+n_j\bb_j},
\label{defI}\end{equation}
where $N_j$ is defined in (\ref{sumK}).
\end{Isum}\goodbreak

\noindent We find that the following identities hold\goodbreak
\newtheorem{IdIs}{Lemma}
\begin{IdIs}
For $\mu_1+\cdots+\mu_L=\ell N+Q$, we find
\begin{enumerate}
\item{ if $\lambda_1+\cdots+\lambda_L=Q$, then
\begin{equation}
I_{kN}(\{\mu_j\};\{\lambda_j\})=\bino{\ell}{k};
\label{IdIs1}\end{equation}}
\item{if $\lambda_1+\cdots+\lambda_L=nN+Q$, then
\begin{equation}
 I_{N}(\{\mu_j\};\{\lambda_j\})=
(\ell-n)+\bI_N(\{\lambda_j\};\{\mu_j\}),
\label{IdIs2}\end{equation}
\begin{equation}
 I_{\ell N}(\{\mu_j\};\{\lambda_j\})=
\bI_{nN}(\{\lambda_j\};\{\mu_j\}),
\label{IdIs3}\end{equation}
\begin{equation}
\fl I_{\ell N-N}(\{\mu_j\};\{\lambda_j\})=
(\ell-n)\bI_{nN}(\{\lambda_j\};\{\mu_j\})+
\bI_{nN-N}(\{\lambda_j\};\{\mu_j\}),
\label{IdIs4}\end{equation}
where
\begin{equation}
\fl\bI_m(\{\lambda_j\};\{\mu_j\})\!=
\sum_{ {\{0\le n_j\le N-1\}}\atop{n_1+\cdots+n_L=m}}
\prod_{j=1}^L\sfactor{\lambda_j}{n_j}\sfactor{n_j+\mu_j}{n_j}\omega^{n_j(\bb_j-\bN_j)+n_j a_j};
\label{defbI}\end{equation}}
\end{enumerate}
\end{IdIs}\goodbreak
\begin{proof} 
First, it is trivial to show $\sum_j n_jN_j=\sum_j n_j\bN_j$,
so that for ${n_1+\cdots+n_L=m}$
$$\sum_{j=1}^L n_jN_j={\frac 12}\sum_{j=1}^L n_j(N_j+\bN_j)=
\frac 12\sum_{j=1}^L n_j(m-n_j)
=\frac 12\Bigg(m^2-\sum_{j=1}^L n_j^2\Bigg).$$
Substituting this into (\ref{defI}) and using (\ref{binomial}) and
(\ref{id1}), we rewrite (\ref{defI}) as
\begin{equation}
\fl(-1)^m\omega^{\halfs m^2}I_m(\{\mu_j\};\{\lambda_j\})\!=
\sum_{ {\{0\le n_j\le N-1\}}\atop{n_1+\cdots+n_L=m}}
\prod_{j=1}^L\frac{(\omega^{-\mu_j};\omega)_{n_j}
(\omega^{1+\lambda_j};\omega)_{n_j}}
{(\omega;\omega)_{n_j}(\omega;\omega)_{n_j}}
\omega^{n_j(\halfs+\mu_j+a_j+\bb_j)}.
\label{defI2}\end{equation}
Its generating function can be written as a product of $L$ sums
\begin{eqnarray}
\prod_{j=1}^L&J_j(t)&=\sum_{m=0}^{a_{L+1}}
(-1)^m\omega^{\halfs m^2}I_m(\{\mu_j\};\{\lambda_j\})t^m,
\quad a_{L+1}=\mu_1+\cdots+\mu_L;\cr
&J_j(t)&=\sum_{n_j=0}^{\mu_j}\frac{(\omega^{-\mu_j};\omega)_{n_j}
(\omega^{1+\lambda_j};\omega)_{n_j}}
{(\omega;\omega)_{n_j}(\omega;\omega)_{n_j}}
(t\omega^{\halfs+\mu_j+a_j+\bb_j})^{n_j}
\label{genI}\end{eqnarray}
where the polynomial $J_j(t)$ is a basic hypergeometric function. The
transformation formula  (10.10.2) of \cite{AnAsRoy} can be rewritten
here as
\begin{equation}
\hypp{2}{1}{q^a,q^b}{q^c}{q,x}=(q^{a+b-c}x;q)^{\vp}_{c-b-a}\;
\hypp{2}{1}{q^{c-a},q^{c-b}}{q^c}{q,q^{a+b-c}x}.
\label{AAR}\end{equation}
Then, setting $q=r\omega$ and letting $r\!\uparrow\!1$ in (\ref{AAR}),
we obtain
\begin{eqnarray}
&J_j(t)&=\hypp{2}{1}{\omega^{-\mu_j},\omega^{1+\lambda_j}}{\omega}
{\omega,t\omega^{\halfs+\mu_j+a_j+\bb_j}}\nonumber\\
&&=(\omega^{\halfs+\lambda_j+a_j+\bb_j}t;
\omega)^{\vp}_{\mu_j-\lambda_j}\;
\hypp{2}{1}{\omega^{1+\mu_j},\omega^{-\lambda_j}}
{\omega}{\omega,t\omega^{\halfs+\lambda_j+a_j+\bb_j}}.
\label{JbJ}\end{eqnarray}
From (\ref{ab}), we find that
\begin{eqnarray}
&&a_1=0,\quad a_j+\mu_j=a_{j+1},\quad a_{L+1}=\ell N+Q, \nonumber\\
&&\bb_0=\lambda_1+\cdots+\lambda_L=nN+Q,\quad
\bb_j+\lambda_j=\bb_{j-1},
\quad\bb_L=0,\label{ab2}\\
&&(\lambda_{j+1}+a_{j+1}+\bb_{j+1})-(\lambda_j+a_j+\bb_j)
=\mu_j-\lambda_j.
\label{ab3}\end{eqnarray}
Using (\ref{ab2}), (\ref{ab3}) and the second identity in (\ref{id2}),
we find
\begin{eqnarray}
&\prod_{j=1}^L
(\omega^{\halfs+\lambda_j+a_j+\bb_j}t;\omega)_{\mu_j-\lambda_j}&=
(\omega^{\halfs+\bb_0}t;\omega)^{\vphantom{P}}_{a_{L+1}-\bb_0}
\nonumber\\
&&=
(\omega^{\halfs+\bb_0}t;\omega)^{\vp}_{\ell N-nN}=(1+t^N)^{\ell-n}.
\label{id3}\end{eqnarray}
The generating function of the sums (\ref{defbI}) is given by
\begin{eqnarray}
\prod_{j=1}^L&\bJ_j(t)&=\sum_{m=0}^{\bb_{0}}(-1)^m\omega^{\halfs m^2}
\bI_m(\{\lambda_j\};\{\mu_j\})t^m,\nonumber\\
&\bJ_j(t)&=\hypp{2}{1}{\omega^{1+\mu_j},\omega^{-\lambda_j}}
{\omega}{\omega,t\omega^{\halfs+\lambda_j+a_j+\bb_j}},
\label{genbI}\end{eqnarray}
in analogy with (\ref{genI}).
Then using (\ref{JbJ}), (\ref{id3}) and (\ref{genbI}), we find
\begin{equation}
\prod_{j=1}^L J_j(t)=(1+t^N)^{\ell-n}\prod_{j=1}^L\bJ_j(t).
\label{gIgbI}\end{equation}
By equating the coefficients of $t^m$ on both sides of (\ref{gIgbI}), we
relate the sums $I_m$ and $\bI_m$.  Particularly, for $\bb_0=Q$, $n=0$,
we can see from (\ref{genbI}) that the generating function of $\bI_m$
is a polynomial of order $Q$ and that the coefficient of $t^{kN}$ can 
only come from the factor in front. Furthermore, as
$(-1)^{kN}\omega^{\halfs (kN)^2}=1$, this proves (\ref{IdIs1}).
For $n\ne0$, by equating the coefficients of $t^N$ in (\ref{gIgbI}),
we prove (\ref{IdIs2}). Finally, (\ref{IdIs3}) and (\ref{IdIs4})
are proved equating the coefficients
of $t^{\ell N}$ and $t^{\ell N-N}$ in (\ref{gIgbI}).
\end{proof}\goodbreak
As a consequence of Lemma 1, we can prove the following identities:
\goodbreak
\newtheorem{kbk}[IdIs]{Lemma}
\begin{kbk}
The sums $K_m(\{n_j\})$ and $\bK_m(\{n_j\})$ defined in (\ref{sumK}) and (\ref{sumbK}) satisfy the following relations
\begin{enumerate}
\item 
\begin{equation}
\fl\sum_{ {\{0\le n_j\le N-1\}}\atop{n_1+\cdots+n_L=kN}}
\bK_{\ell N+Q}(\{n_j\})K_Q(\{n_j\})
=\bino{\ell+k}{k}\Lambda^{Q}_{0}\Lambda^{Q}_{\ell+k},\quad
\Lambda^Q_m=c_{mN+Q};
\label{Idkbk1}\end{equation}
\item
\begin{equation}
\fl\sum_{ {\{0\le n_j\le N-1\}}\atop{n_1+\cdots+n_L=N}}
\bK_{\ell N+Q}(\{n_j\})K_{m N+Q}(\{n_j\})
=\sum_{n=0}^m(\ell+1+2n-m)\Lambda^{Q}_{m-n} \Lambda^{Q}_{\ell+1+n}.
\label{Idkbk2}\end{equation}
\end{enumerate}
\end{kbk}\goodbreak
\begin{proof} 
Denote
\begin{equation}
\Theta_{\ell,m,k}\equiv
\sum_{ {\{0\le n_j\le N-1\}}\atop{n_1+\cdots+n_L=kN}}
\bK_{\ell N+Q}(\{n_j\})K_{mN+Q}(\{n_j\}).
\label{theta}\end{equation}
Using (\ref{sumK}) and (\ref{sumbK}), we find
\begin{equation}
\Theta_{\ell,m,k}=
\sum_{ {\{0\le \lambda_j,n_j,n'_j\le N-1\}}\atop{n_1+\cdots+n_L=kN}}
\prod_{j=1}^L\sfactor{n'_j+n_j}{n_j}
\sfactor{n_j+\lambda_j}{n_j}\omega^{n_j(N'_j+\bb_j)},
\label{lhs}\end{equation}
where 
\begin{eqnarray}
N'_j&=\sum_{i<j}n'_i,\quad
N'_{L+1}&=n'_1+\cdots+n'_L=\ell N+Q,\nonumber\\
\bb_j&=\sum_{i>j}\lambda_j,\quad\phantom{N'}
\bb_0&=\lambda_1+\cdots+\lambda_L=mN+Q.
\end{eqnarray}
In order to apply (\ref{defI}) we change the $L$ summation variables
$n'_j$ to $\mu_j=n_j+n'_j$. Since $n_j<N$, we find that the summand
in (\ref{lhs}) is nonzero if and only if $\mu_j<N$. Also,
$a_{L+1}=\mu_1+\cdots+\mu_L=(\ell+k)N+Q$.
Thus we may rewrite (\ref{lhs}) as
\begin{equation}
\Theta_{\ell,m,k}=
\sum_{{\{0\le \mu_j\le N-1\}}\atop{\mu_1+\cdots+\mu_L=(\ell+k)N+Q}}
\sum_{{\{0\le\lambda_j\le N-1\}}\atop{\lambda_1+\cdots+\lambda_L=mN+Q}}
I_{kN}(\{\mu_j\};\{\lambda_j\}).
\label{lhs2}\end{equation}
Now we let $m=0$ and using (\ref{IdIs1}) with $\ell$ replaced
by $\ell+k$ we find
\begin{equation}
\Theta_{\ell,0,k}=\sum_{ {\{0\le \mu_j\le N-1\}}\atop{\mu_1+\cdots+\mu_L=(\ell+k)N+Q}}
\sum_{ {\{0\le \lambda_j\le N-1\}}\atop{\lambda_1+\cdots+\lambda_L=Q}}
\bino{\ell+k}{k}=\bino{\ell+k}{k}\Lambda_{\ell+k}^Q\Lambda_0^Q,
\label{lhs3}\end{equation}
where $K_{mN+Q}(\{0\})=c_{mN+Q}=\Lambda^Q_m$ is used. This proves (\ref{Idkbk1}).

Choosing $k=1$ in (\ref{lhs2}) and using (\ref{IdIs2}) with $\ell$
replaced by $\ell+1$, we find
\begin{eqnarray}
&\fl\Theta_{\ell,m,1}&=(\ell+1-m)\Lambda_{\ell+1}^Q\Lambda_m^Q
+\sum_{{\{0\le\mu_j\le N-1\}}\atop{\mu_1+\cdots+\mu_L=(\ell+1)N+Q}}
\sum_{{\{0\le\lambda_j\le N-1\}}\atop{\lambda_1+\cdots+\lambda_L=mN+Q}}
\bI_{N}(\{\lambda_j\},\{\mu_j\})\nonumber\\
&\fl&=(\ell+1-m)\Lambda_{\ell+1}^Q\Lambda_m^Q+\Theta_{\ell+1,m-1,1},
\label{lhs4}\end{eqnarray}
where we have used (\ref{lhs}) with $k=1$ after first replacing
$\lambda_j$ by $\lambda_j+n_j$ and $\mu_j$ by $n'_j$ in (\ref{defbI}).
Note that summands with $\lambda_j<n_j$ in (\ref{defbI}) or with
$\lambda_j+n_j>N$ in (\ref{lhs}) vanish.
We next apply (\ref{lhs4}) to $\Theta_{\ell+1,m-1,1}$ and repeat the 
process until arriving at $\Theta_{\ell+m,0,1}$. Then we can use (\ref{Idkbk1}) with $k=1$ to obtain (\ref{Idkbk2}). 
\end{proof}
\noindent{\em Remark.}
Substituting $n=m-j$ we may rewrite (\ref{Idkbk2}) as
\begin{equation}
\Theta_{\ell,m,1}
=\sum_{j=0}^m(\ell+1+m-2j)\Lambda^{Q}_{j} \Lambda^{Q}_{\ell+1+m-j}.
\label{Idkbk3}\end{equation}

Now we are ready to prove the orthogonality theorem.
\begin{proof}
As in (II.72) we introduce the polynomial
\begin{equation}
{\mbox{\myeu h}}^Q_k(z)\equiv
\sum_{ {\{0\le n_j\le N-1\}}\atop{n_1+\cdots+n_L=N}}
\bG_Q(\{n_j\},z_{k,Q})G_Q(\{n_j\},z).
\label{dh}\end{equation}
Substituting (\ref{GbG1}) with $k=1$ into this equation, we find
\begin{equation}
\fl{\mbox{\myeu h}}^Q_k(z_{i,Q})=\sum_{\ell=0}^{m_Q-1}\sum_{m=0}^{m_Q-1}z_{k,Q}^\ell z_{i,Q}^m
\sum_{ {\{0\le n_j\le N-1\}}\atop{n_1+\cdots+n_L=N}}
\bK_{\ell N+Q}(\{n_j\})K_{mN+Q}(\{n_j\}).
\label{hki}\end{equation}
Now we use Lemma~2({\em ii}) or (\ref{Idkbk3}) to write
\begin{equation}
\fl{\mbox{\myeu h}}^Q_k(z_{i,Q})=\sum_{\ell=0}^{m_Q-1}\sum_{m=0}^{m_Q-1}z_{k,Q}^\ell z_{i,Q}^m
\sum_{j=0}^m
(\ell+m+1-2j)\Lambda^{Q}_{j} \Lambda^{Q}_{\ell+m+1-j}.
\label{hki1}\end{equation}
Interchanging the order of summation over $m$ and $j$, and then letting
$m=n+j-\ell-1$, we find
\begin{equation}
{\mbox{\myeu h}}^Q_k(z_{i,Q})=\sum_{\ell=0}^{m_Q-1}\sum_{j=0}^{m_Q-1}
\sum_{n=\ell+1}^{m_Q+\ell-j}z_{k,Q}^\ell z_{i,Q}^{n+j-\ell-1}
(n-j)\Lambda^{Q}_{j} \Lambda^{Q}_{n}.
\label{hki1a}\end{equation}
Since $\Lambda^Q_n=0$ for $n>m_Q$, the intervals of summation may be extended to $0\le\ell,j\le m_Q$.

We split the summation over $n$ into three parts as\footnote{Note
that it follows from (\ref{Q}) and (\ref{DrinP}) that none of
the $z_{i,Q}$ vanishes.}
\begin{equation}
\fl{\mbox{\myeu h}}^Q_k(z_{i,Q})=\sum_{\ell=0}^{m_Q}
\Bigl(\frac{z_{k,Q}}{z_{i,Q}}\Bigr)^\ell\sum_{j=0}^{m_Q}
\Biggl[\sum_{n=0}^{m_Q}-\sum_{n=0}^{\ell}-
\sum_{n=m_Q+\ell-j+1}^{m_Q}\Biggr]
(n-j)\Lambda^{Q}_{j} \Lambda^{Q}_{n}z_{i,Q}^{n+j-1}.
\label{hki2}\end{equation}
The contribution due to the first part is identically zero, as it is
antisymmetric under the interchange of the summation variables $j$
and $n$. The terms with $j\le \ell$ of the third part are zero, as
$\Lambda^Q_n=0$ for $n>m_Q$, leaving the nontrivial terms
\begin{eqnarray}
\fl\sum_{j=\ell+1}^{m_Q}\;\sum_{n=m_Q+\ell-j+1}^{m_Q}(n-j)
\Lambda^{Q}_{j} \Lambda^{Q}_{n}z_{i,Q}^{n+j-1}&&=
\sum_{n=\ell+1}^{m_Q}\:\sum_{j=m_Q+\ell-n+1}^{m_Q}(n-j)
\Lambda^{Q}_{j} \Lambda^{Q}_{n}z_{i,Q}^{n+j-1}\cr
&&=\sum_{j=\ell+1}^{m_Q}\:\sum_{n=m_Q+\ell-j+1}^{m_Q}(j-n)
\Lambda^{Q}_{j} \Lambda^{Q}_{n}z_{i,Q}^{n+j-1}=0.
\end{eqnarray}
Here the first equality is obtained by interchanging the order of the summations, the second by interchanging $n$ and $j$. This shows
that the third part is also identically zero.

We split the second part into two pieces with summands proportional
to $n$ and to $j$. In the first piece we can perform the sum over $j$,
which by (\ref{roots}) yields a factor $P_Q(z_{i,Q})$. This shows that
the only nonvanishing contribution comes from the second piece, or
\begin{equation}
\fl{\mbox{\myeu h}}^Q_k(z_{i,Q})=
\sum_{\ell=0}^{m_Q}\Bigl(\frac{z_{k,Q}}{z_{i,Q}}\Bigr)^\ell
\sum_{j=0}^{m_Q}\sum_{n=0}^{\ell}
j\Lambda^{Q}_{j} \Lambda^{Q}_{n}z_{i,Q}^{n+j-1}=
\sum_{j=0}^{m_Q}\sum_{n=0}^{m_Q}j\Lambda^{Q}_{j} 
\Lambda^{Q}_{n}z_{i,Q}^{n+j-1}
\sum_{\ell=n}^{m_Q}\Bigl(\frac{z_{k,Q}}{z_{i,Q}}\Bigr)^\ell.
\label{hki3}\end{equation}
We first consider the case $z_{k,Q}\ne z_{i,Q}$, so that the last
sum can be evaluated to give
\begin{equation}
\fl{\mbox{\myeu h}}^Q_k(z_{i,Q})=\frac1{1-(z_{k,Q}/z_{i,Q})}
\sum_{j=0}^{m_Q}\sum_{n=0}^{m_Q}j\Lambda^{Q}_{j}
\Lambda^{Q}_{n}z_{i,Q}^{n+j-1}\Biggl[
\biggl(\frac{z_{k,Q}}{z_{i,Q}}\biggr)^n-
(\frac{z_{k,Q}}{z_{i,Q}}\biggr)^{m_Q+1}\Biggr]=0,
\label{hki4}\end{equation}
as seen again from (\ref{roots}) after summing over $n$.
Finally, we consider the case  $z_{k,Q}= z_{i,Q}$, so that the
last sum is $(m_Q-n+1)$ leading to
\begin{equation}
\fl{\mbox{\myeu h}}^Q_k(z_{k,Q})=\sum_{j=0}^{m_Q}
\sum_{n=0}^{m_Q}j\Lambda^{Q}_{j}
\Lambda^{Q}_{n}z_{k,Q}^{n+j-1}(m_Q-n+1)
=-z_{k,Q}\biggl[\sum_{j=0}^{m_Q} j
\Lambda^{Q}_{j} z_{k,Q}^{j-1}\biggr]^2,
\label{hki5}\end{equation}
after using (\ref{roots}) to show that the $(m_Q+1)$ terms do not contribute. We use (\ref{roots}) once more to show
\begin{equation}
\fl P'_Q(z)=\sum_{j=0}^{m_Q} j \Lambda^{Q}_{j} z^{j-1}
=\Lambda^{Q}_{ m_Q}\left(\,\prod_{j\ne k}(z-z_{j,Q})+
(z-z_{k,Q})\frac{\rmd}{\rmd z}\prod_{j\ne k}(z-z_{j,Q})\right).
\label{hki6}\end{equation}
Consequently, we have 
\begin{equation}
\fl\sum_{j=0}^{m_Q} j \Lambda^{Q}_{j} z_{k,Q}^{j-1}
=\Lambda^{Q}_{ m_Q}\prod_{j\ne k}(z_{k,Q}-z_{j,Q}), \quad
{\mbox{\myeu h}}^Q_k(z_{k,Q})
=z_{k,Q}(\Lambda^{Q}_{ m_Q})^2\prod_{j\ne k}(z_{k,Q}-z_{j,Q})^2.
\label{hki7}\end{equation}
This is identical to $B_k$ in (\ref{conB}). Thus we have finished
the proof.
\end{proof}

We note that we have not used the assumption that all roots $z_{i,Q}$
for given $L$ and $Q$ are distinct. The theorem holds also, if there
is a multiple root $z_{k,Q}$, for which $B_k=0$. However, there is
substantial numerical evidence that the roots may all be distinct and
we shall assume this for the following corollary of the theorem.

The degree of polynomial (\ref{dh}) is $m_Q-1$, as can be seen
from (\ref{GbG1}) with $k=1$, whereas polynomial
(II.10) \cite{APsu2,Davies}
\begin{equation}
f^Q_k(z)=\prod_{\ell\ne k}{\frac{z-z_{\ell,Q}}{z_{k,Q}-z_{\ell,Q}}}=
\sum_{n=0}^{m_Q-1}\beta^Q_{k,n}z^n, \quad
f^Q_k(z_{j,Q})=\delta_{j,k}
\label{beta}\end{equation}
is of the same degree and has the same roots. These
polynomials must be equal except for some multiplicative constant.
Using the orthogonality theorem, we find the identity
\begin{equation}
{\mbox{\myeu h}}^Q_k(z)=-B_k f^Q_k(z)=
-z_{k,Q}(\Lambda^{Q}_{ m_Q})^2
\prod_{\ell\ne k}\,[(z_{k,Q}-z_{\ell,Q})(z-z_{\ell,Q})],
\label{coro}\end{equation}
valid when all roots are distinct.

Finally, we can also introduce the polynomials
\begin{equation}
\bar{\mbox{\myeu h}}^Q_k(z)\equiv
\sum_{ {\{0\le n_j\le N-1\}}\atop{n_1+\cdots+n_L=N}}
G_Q(\{n_j\},z_{k,Q})\bG_Q(\{n_j\},z).
\label{corb}\end{equation}
Since the roots $z_{k,Q}$ are real, $\bar{\mbox{\myeu h}}^Q_k(z)$ is
the complex conjugate of ${\mbox{\myeu h}}^Q_k(z)$ for real $z$. But by
(\ref{coro}) we see that ${\mbox{\myeu h}}^Q_k(z)$ is real in that case,
so that
\begin{equation}
\bar{\mbox{\myeu h}}^Q_k(z)=
{\mbox{\myeu h}}^Q_k(z),\quad\hbox{if } z\in\mathbb{R}.
\label{corob}\end{equation}

\section*{Acknowledgments}
We thank our colleagues and the staff at the Centre for Mathematics and
its Applications (CMA) and at the Department of Theoretical Physics
(RSPE) of Australian National University for their generous support and
hospitality.

\end{document}